# Pulmonary electrical impedance tomography based on deep recurrent neural networks

Zhenzhong Song[1, 2*], Jianping Li [2], Jun Zhang[1], Hanyun Wen[1], Suqin Zhang[1], Wei Jiang[1] and Xingxing Zhou[1]

[1]Shizhen College of Guizhou University of Traditional Chinese Medicine, Guiyang, Guizhou, China;  
[2]Zhejiang Normal University, Jinhua, Zhejian, China  
E-mail: sphil12@zjnu.edu.cn

**Abstract**

Electrical impedance tomography (EIT) is a non-invasive functional imaging technology. In order to enhance the quality of lung EIT images, novel algorithms, namely LSTM-LSTM, LSTM-BiLSTM, BiLSTM-LSTM, and BiLSTM-BiLSTM, leveraging LSTM or BiLSTM networks, were developed. Simulation results demonstrate that the optimized deep recurrent neural network significantly enhanced the quality of the reconstructed images. Specifically, the correlation coefficients of the LSTM-LSTM and the LSTM-BiLSTM algorithms exhibited maximum increases of 27.5% and 25.4% over the LSTM algorithm, respectively. Moreover, in comparison to the BiLSTM algorithm, the correlation coefficients of the BiLSTM-LSTM and BiLSTM-BiLSTM algorithms increased by 11.7% and 13.4%, respectively. Overall, the quality of EIT images showed notable enhancement. This research offers a valuable approach for enhancing EIT image quality and presents a novel application of LSTM networks in EIT technology.

## 1. Introduction

Currently, acute respiratory distress syndrome (ARDS), chronic obstructive pulmonary disease (COPD), and various other lung diseases pose significant threats to human life. The advancement of monitoring technologies for lung diseases is ongoing, with medical imaging emerging as a primary modality for qualitative and quantitative assessment of lung function. Common modalities include CT imaging, ultrasound imaging, and electrical impedance tomography (EIT). EIT, a novel medical diagnostic technology, offers attributes such as safety, non-invasiveness, non-radiation, and real-time monitoring capabilities(Onsager et al 2024, Rubin et al 2022, Maciejewski et al 2021). Consequently, EIT finds extensive application in clinical medical monitoring, material detection(Thomas et al 2019, Kuusela et al 2025, Gupta et al 2021), and other domains. In clinical settings, EIT is predominantly utilized for monitoring brain diseases (Bai et al 2024, Li et al 2025, Shi et al 2024), breast diseases(Ouypornkochagorn et al 2024, Setyawan et al 2024, Jin et al 2023), and pulmonary diseases (Gaulton et al 2023, Bayford et al 2024, Martin et al 2023, Brito et al 2023, Wu et al 2025), showing promising potential.

The efficacy of EIT technology relies significantly on the quality of reconstructed images. Higher quality images yield more accurate information, while lower quality images may lead to increased errors and potential misinterpretations. Enhancing the quality of reconstructed images is crucial for advancing the application of EIT. This improvement is going to be achieved through two main approaches. Firstly, research efforts are directed towards enhancing the precision of measurements and minimizing interference in the hardware components of the imaging equipment.

From a software standpoint, the primary objective is to enhance the robustness and precision of the reconstruction algorithms. Typically, optimization techniques are employed to refine the structure and computational efficiency of the algorithms, thereby enhancing the quality of the resulting images.

Various reconstruction algorithms are utilized in the field. These algorithms encompass convolutional neural networks (CNN)(Okamura *et al* 2024,Lou *et al* 2025), BP neural networks(Song *et al* 2024), radial basis neural networks(Qian *et al* 2024, Dong *et al* 2023), Gauss-Newton iterative algorithms(Shin *et al* 2024), and the GVSPM algorithm(Dong *et al* 2004). Long short-term memory (LSTM), a neural network architecture, effectively addresses the challenges of gradient vanishing or exploding gradients and captures the prolonged dependencies within data(Klosowski et al 2023a 2023b). Consequently, in this research, EIT images are reconstructed employing the LSTM algorithm, with a focus on enhancing the quality of the reconstructed images through optimizing the LSTM algorithm.

## 2 Principle of electrical impedance tomography

Electrical impedance tomography is a diagnostic technique that utilizes conductivity distribution to assess internal features or changes within an object. Initially, electrodes are affixed to the object's surface. Subsequently, an excitation signal is introduced into the object via two electrodes. Then the resulting induced potential on other electrodes is measured to determine the voltage between the two electrodes. This measured voltage is utilized to reconstruct the internal conductivity distribution of the object. The EIT system, illustrated in Figure 1, comprises electrodes, a data acquisition unit, and additional components. A computer (PC) directs the data acquisition unit to deliver the excitation signal to the tissue surface of the object through a pair of electrodes, while the induced voltage is captured by the remaining electrode. Subsequently, the data acquisition unit relays the voltage data back to the PC, where a reconstruction algorithm is employed to generate EIT images.

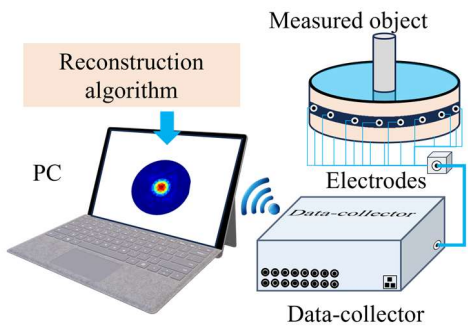

Fig. 1 Diagram of the electrical impedance tomography system

The solution process of the EIT technology comprises two primary stages: solving the forward problem and solving the inverse problem. The forward problem involves determining the boundary voltage given the excitation signal and known conductivity distribution. Conversely, the inverse problem entails measuring the boundary voltage with a known excitation signal to deduce the unknown conductivity distribution through reconstruction algorithms, as depicted in Figure 2.

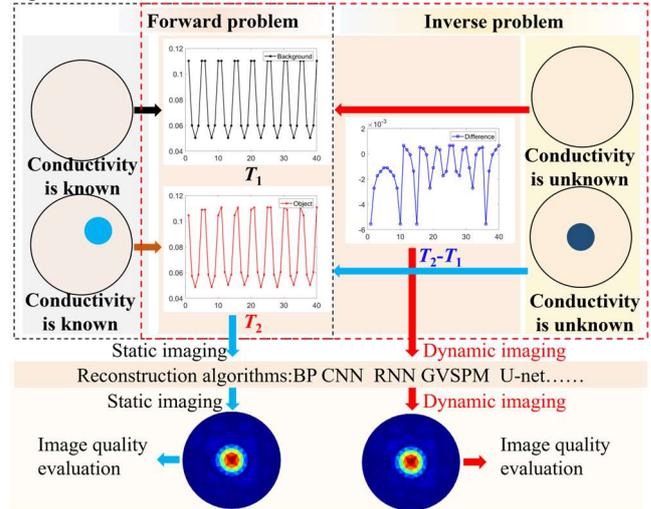

Fig. 2 Principle of electrical impedance tomography

The method of reconstructing images is categorized into static imaging and dynamic imaging [29]. Static imaging involves reconstructing EIT images of objects by applying boundary voltages at specific moments, resulting in direct image reconstruction using the boundary voltage at time T2. On the other hand, dynamic imaging reconstructs EIT images by utilizing the differences in boundary voltages collected at different time points, with image reconstruction based on the boundary voltage variances between times T2 and T1. Dynamic imaging is advantageous as it is less susceptible to noise interference, thereby ensuring stable image quality. In contrast, static imaging is characterized by its speed, although it necessitates high accuracy in measurement equipment. In this study, dynamic imaging and adjacent excitation measurements were employed to acquire boundary voltage data. Specifically, an 8-electrode EIT system was utilized to gather 40 boundary voltage values upon completion of a single excitation measurement.

## 3 Deep recurrent neural networks

Developing an advanced reconstruction algorithm is essential for enhancing the quality of EIT images. Optimization of the reconstruction algorithm plays a crucial role in improving the quality of reconstructed images. In this study, long short-term memory is employed to enhance the reconstruction of EIT images, with its architecture tailored and optimized to achieve superior image quality.

## 3.1 Recurrent neural network

LSTM represents a specialized type of recurrent neural network (RNN) adept at capturing temporal dynamics within datasets, particularly suited for sequences. Unlike conventional fully connected neural networks, where neuron states in each layer rely solely on those from the preceding layer. LSTM allows for dependencies across time steps, crucial for modeling sequential data. This capability to retain and utilize past neuron states enables LSTM to effectively extract intricate features from sequential datasets. It is unattainable for traditional fully connected neural networks.

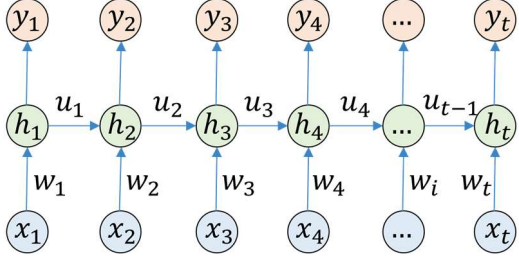

Fig. 3 The structure of RNN

Figure 3 illustrates the basic architecture of a simple recurrent neural network, which comprises an input layer, a hidden layer, and an output layer. In contrast to a fully connected network, the hidden layer neurons in an RNN interact with each other. The state of a neuron in the hidden layer, denoted as $h_t$, can be expressed as

$$h_t = f(w_t x_t + h_{t-1} u_{t-1} + b_t) \quad (1)$$

here, $f(\cdot)$ represents the activation function, and $W=(w_1, w_2, w_3,... w_t)$ indicates the state-output weight matrix. $U=(u_1, u_2, u_3,... u_t)$ represents the state-state weight matrix, and $B=(b_1, b_2, b_3,... b_t)$ represents the bias matrix.

## 3.2 Long short-term memory network

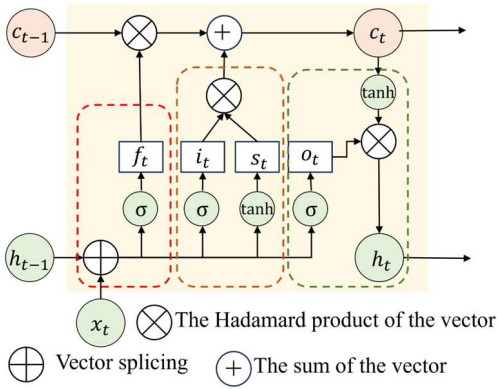

Fig. 4 The structure of LSTM

The LSTM network effectively addresses the issue of gradient vanishing or exploding encountered in traditional RNN networks. In comparison to conventional RNNs, LSTM incorporates additional components such as a forgetting gate ($f_t$), an input gate ($i_t$), an output gate ($o_t$), and an internal state ($c_t$), as illustrated in Figure 4. The forgetting gate regulates the retention of information from the previous internal state ($c_{t-1}$). The input gate manages the integration of memory information from the candidate state ($\hat{c}_t$) at the current time step. The output gate governs the flow of information between the internal state ($c_t$) and the external state ($h_t$).

The $f_t$, $i_t$, and $o_t$ are calculated by using the previous external state $h_{t-1}$ and the current input $x_t$. The calculation formula is as follows:

$$i_t = \sigma(w_i x_i + u_i h_{t-1} + b_i) \quad (2)$$

$$f_t = \sigma(w_f x_i + u_f h_{t-1} + b_f) \quad (3)$$

$$o_t = \sigma(w_o x_i + u_o h_{t-1} + b_o) \quad (4)$$

here, $\sigma(.)$ represents the activation function. The states of the input gate, output gate, and forget gate are denoted by 0 or 1, respectively. If the state is 0, that indicates the closed state and blocking information. It indicates the open state and allows information to pass through under the condition that the state is 1. Thus, $f_t \in [0,1]^D$, $i_t \in [0,1]^D$, and $o_t \in [0,1]^D$.

After the state information of the three gates is determined, the calculating formula for the candidate state $\hat{c}_t$ is as follows:

$$\hat{c}_t = \tanh(w_c x_t + u_c h_{t-1} + b_c) \quad (1.1)$$

At the moment, the internal state $c_t$ and the external state $h_t$ are calculated as follows, respectively.

$$c_t = f_t \odot c_{t-1} + i_t \odot \hat{c}_t \quad (1.2)$$

$$h_t = o_t \odot \tanh(c_t) \quad (1.3)$$

here, $\odot$ represents the Hadamard product of vectors.

## 3.3 Bidirectional long short-term memory network

Bidirectional long short-term memory (BiLSTM) comprises two layers of recurrent neural networks with identical inputs. The distinction lies in the direction of information flow, as illustrated in Figure 5. While the first layer transmits information sequentially, the second layer does so in reverse. The computational procedure mirrors that of the LSTM network.

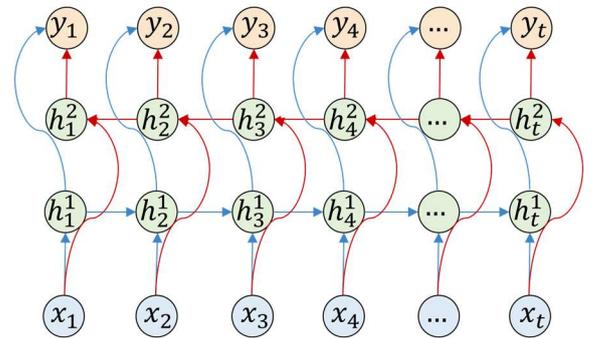

Fig. 5 The structure of BiLSTM

## 3.4 Deep recurrent neural network

A deep recurrent neural network is an advanced network architecture achieved by incorporating multiple layers of neurons into the existing RNN framework. The bidirectional recurrent neural network, another variant of the deep recurrent neural network, differs in its design. This research introduces a novel deep recurrent neural network by integrating LSTM and BiLSTM networks, departing from conventional deep recurrent neural network structures.

The BiLSTM network is constructed based on LSTM networks, resulting in the LSTM-BiLSTM algorithm. Conversely, the LSTM network is built upon BiLSTM networks to create the BiLSTM-LSTM algorithm, as illustrated in Fig. 6. A comparative analysis of the post-stack reconstruction algorithm involves evaluating the outputs $R_i$ ($i$=1,2,3,4) of the LSTM network, BiLSTM network, LSTM-BiLSTM algorithm, and BiLSTM-LSTM algorithm, along with a comparison of the quality of reconstructed images. Furthermore, the LSTM-LSTM algorithm is established when the LSTM network is stacked on another LSTM network. Similarly, the BiLSTM-BiLSTM algorithm is developed, as depicted in Fig. 7.

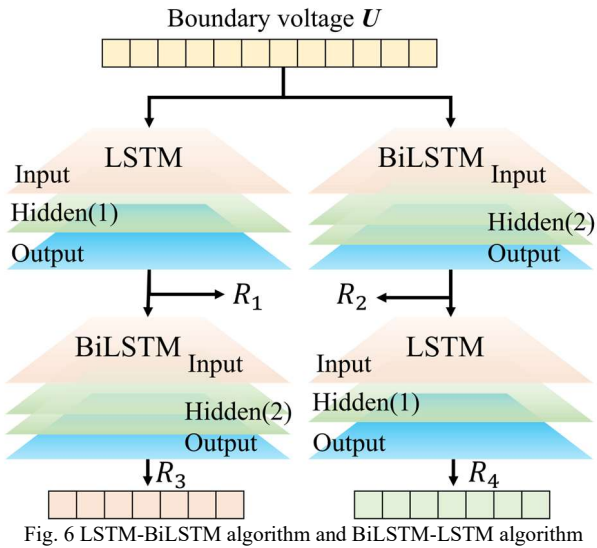

Fig. 6 LSTM-BiLSTM algorithm and BiLSTM-LSTM algorithm

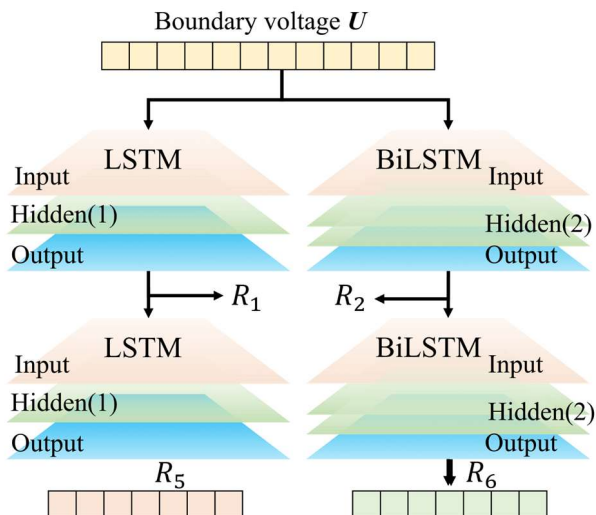

Fig. 7 LSTM-LSTM algorithm and BiLSTM-BiLSTM algorithm

## 4 Finite Element Analysis

### 4.1 Finite Element Models

Finite element analysis (FEA) is commonly employed to discretize continuous problems. Various finite element distributions are utilized to approximate the conductivity distribution of the measured field, as illustrated in Figure 8. The circular cross-section and lung cross-section models are applied to assess the effect of reconstructed images on deep recurrent neural networks. FEA was conducted on both models, each comprising 576 finite elements, corresponding to the number of conductivities being 576.

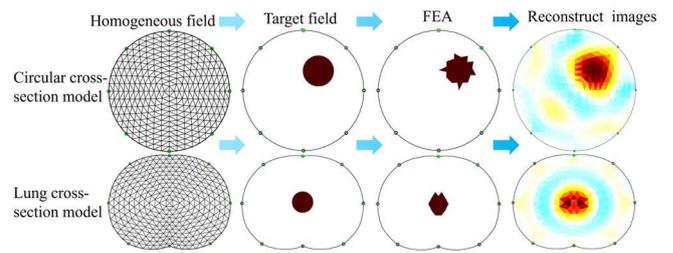

Fig. 8 Finite element analysis of field model

Upon division of the measured field by finite element analysis, discrepancies arise between the contour of the target object and the predetermined contour. As the number of finite elements increases, the target object's contour progressively converges towards the predetermined contour. Nevertheless, this iterative refinement escalates computational complexity and time duration. Consequently, the measured area is discretized into 576 finite elements.

### 4.2 Datasets

The neural network model for reconstructing EIT images requires training with datasets. Consequently, a significant amount of data needs to be acquired. To expand the dataset for circular and lung cross-section models, the constrained condition is systematically modified, as illustrated in Figure 9.

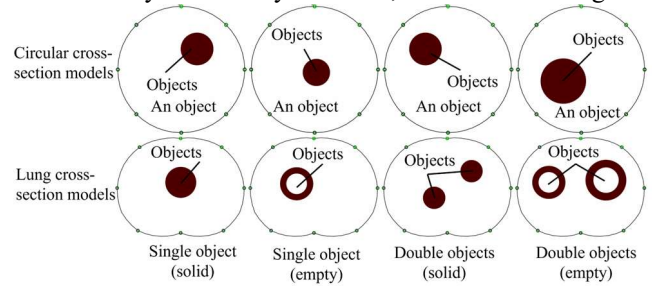

Fig. 9 Finite element analysis of the circular cross-section models and lung cross-section models

These datasets of the circular cross-section models are obtained that vary in the size, position, and conductivity of objects to preliminarily assess the reconstruction algorithms. A total of 300 datasets were gathered for this purpose. On the other hand, lung cross-sectional models involve datasets that vary in the number, size, position, and conductivity of solid and hollow objects. To thoroughly evaluate the reconstruction algorithm, 60000 datasets were collected for this specific purpose.

## 5 Simulation Results

### 5.1 Reconstructing the EIT images

To assess the fidelity of reconstructed images generated by deep recurrent neural networks, these models are selected for simulation analysis about a set of five circular cross-section models and five lung cross-section models. Among them, there is only one object in the five circular cross-section models. In contrast, the lung cross-section models exhibit discrepancies in object size, quantity, position, and shape. Six reconstruction algorithms were employed to generate EIT images for approximately ten models. The simulation outcomes are depicted in Figures 10 and 11.

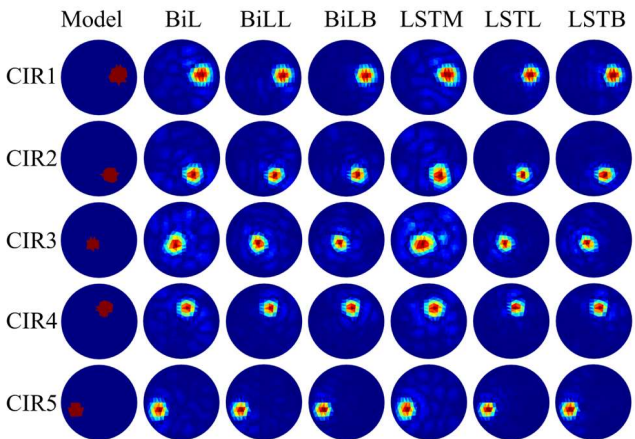

Fig. 10 Simulation results of the circular cross-section models

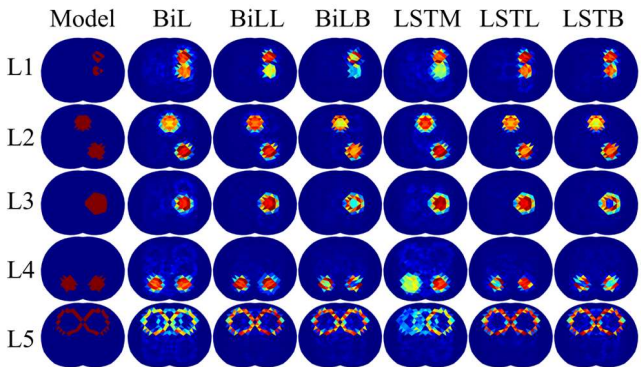

Fig. 11 Simulation results of the lung cross-section models

In Figure 10, the symbols BiL and LSTL represent the BiLSTM algorithm and the LSTM-LSTM algorithm, respectively. The other symbols are described in Table 1.

Table 1 Description of symbols

| Symbol | Meaning |
| --- | --- |
| BiL | BiLSTM algorithm |
| BiLL | BiLSTM-LSTM algorithm |
| BiLB | BiLSTM-BiLSTM algorithm |
| LSTM | LSTM algorithm |
| LSTL | LSTM-LSTM algorithm |
| LSTB | LSTM-BiLSTM algorithm |

The simulation results depicted in Figure 10 reveal an increased presence of artifacts in the reconstructed images generated by both the BiLSTM and LSTM algorithms. It leads to a higher occurrence of blue shadows within the depicted area. This observation underscores the necessity for enhancing the quality of the reconstructed images. Conversely, the optimized image has fewer artifacts and demonstrates superior-quality image.

Five lung cross-section models were utilized to assess the quality of the reconstructed images. The simulation results for the lung cross-section model are illustrated in Figure 11. The analysis reveals that the reconstructed images generated by the BiLSTM and LSTM algorithms exhibit minimal artifacts. However, there exists a considerable shape error in the object contour compared with the actual images. This disparity is most pronounced in the lung cross-sectional model L5. Conversely, the shape error is notably reduced in the BiLSTM-LSTM, BiLSTM-BiLSTM, LSTM-LSTM, and LSTM-BiLSTM algorithms. The reconstructed images closely resemble the actual objects with minor discrepancies. Furthermore, the conductivity distribution in the reconstructed images closely approximates the actual conductivity distribution.

### 5.2 The quality assessment of EIT images

The correlation coefficient (CC) is a crucial parameter for assessing the quality of reconstructed images in digital evaluation(Katayama *et al* 2024, Deng *et al* 2023). A higher correlation coefficient indicates better image quality with fewer artifacts. While a lower coefficient signifies poorer quality, potentially leading to issues like contour distortion. The correlation coefficient is expressed as follows:

$$CC = \frac{\sum_{i=1}^{n}(x_i - x_m)(y_i - y_m)}{\sqrt{\sum_{i=1}^{n}(x_i - x_m)^2}\sqrt{\sum_{i=1}^{n}(y_i - y_m)^2}} \quad (8)$$

here, *CC* denotes the correlation coefficient. *x* and *y* represent the reconstructed conductivity distribution and the actual conductivity distribution, respectively. $x_m$ and $y_m$ represent the average value of *x* and *y*, respectively. The correlation coefficient is utilized to evaluate the quality of the reconstructed images of the circular cross-section and lung cross-section models. The corresponding computational outcomes are detailed in Tables 2 and 3.

Table 2 Correlation coefficient of the circular cross-section models

| CC | CIR1 | CIR2 | CIR3 | CIR4 | CIR5 |
| --- | --- | --- | --- | --- | --- |
| BiL | 0.881 8 | 0.871 9 | 0.728 5 | 0.879 2 | 0.882 6 |
| BiLL | 0.902 3 | 0.885 6 | 0.799 3 | 0.896 9 | 0.907 7 |
| BiLB | 0.908 9 | 0.896 4 | 0.826 2 | 0.894 6 | 0.915 4 |
| LSTM | 0.876 2 | 0.821 5 | 0.665 0 | 0.858 6 | 0.853 6 |
| LSTL | 0.910 3 | 0.897 5 | 0.847 7 | 0.889 1 | 0.915 9 |
| LSTB | 0.903 8 | 0.899 8 | 0.834 1 | 0.898 5 | 0.911 1 |

Tables 2 and 3 demonstrate that in the circular cross-section model, the BiLSTM network outperforms the LSTM network in terms of reconstructed image quality. Conversely, in certain lung cross-section models, the LSTM network produces higher-quality reconstructed images compared to the BiLSTM network. Overall, there is minimal disparity in the correlation coefficients between the two algorithms. Across all models, the correlation coefficients of the LSTM-LSTM and LSTM-BiLSTM algorithms exceed those of the LSTM algorithm, with the exception of models L3 and L4. Notably, the correlation coefficients of the BiLSTM-LSTM and BiLSTM-BiLSTM algorithms show significant improvement compared with the BiLSTM algorithm.

Table 3 Correlation coefficient of the lung cross-section models

| CC | L1 | L2 | L3 | L4 | L5 |
|---|---|---|---|---|---|
| BiL | 0.657 9 | 0.931 3 | 0.843 9 | 0.933 3 | 0.879 7 |
| BiLL | 0.734 9 | 0.963 1 | 0.908 8 | 0.934 8 | 0.950 6 |
| BiLB | 0.679 2 | 0.973 6 | 0.889 2 | 0.889 6 | 0.947 3 |
| LSTM | 0.645 5 | 0.939 0 | 0.918 2 | 0.864 3 | 0.736 5 |
| LSTL | 0.740 0 | 0.969 1 | 0.926 8 | 0.861 1 | 0.955 2 |
| LSTB | 0.756 0 | 0.971 4 | 0.895 5 | 0.894 3 | 0.954 0 |

By comparing the correlation coefficients presented in Tables 2 and 3, it is evident that the stacked networks utilizing either the BiLSTM or LSTM architecture significantly enhance the quality of reconstructed images. The maximum correlation coefficients achieved by the BiLSTM-LSTM and BiLSTM-BiLSTM configurations are 0.9631 and 0.9736, respectively. Similarly, the maximum correlation coefficients for the LSTM-LSTM and LSTM-BiLSTM setups are 0.9691 and 0.9714, respectively. The correlation coefficients among the four algorithms exhibit minimal variation. To facilitate a clearer analysis of the correlation coefficient changes, the increase ratio of the correlation coefficients is computed for pre- and post-algorithm optimization, as detailed in Table 4.

Tab. 4 The increase ratio of correlation coefficient before and after algorithm optimization

| Increase ratio | BiL-BiLL | BiL-BiLB | LSTM-LSTL | LSTM-LSTB |
|---|---|---|---|---|
| CIR1 | 0.023 | 0.031 | 0.039 | 0.032 |
| CIR2 | 0.016 | 0.028 | 0.092 | 0.095 |
| CIR3 | 0.097 | 0.134 | 0.275 | 0.254 |
| CIR4 | 0.020 | 0.018 | 0.036 | 0.047 |
| CIR5 | 0.028 | 0.037 | 0.073 | 0.067 |
| L1 | 0.117 | 0.032 | 0.146 | 0.171 |
| L2 | 0.034 | 0.045 | 0.032 | 0.035 |
| L3 | 0.077 | 0.054 | 0.009 | -0.025 |
| L4 | 0.002 | -0.047 | -0.004 | 0.035 |
| L5 | 0.081 | 0.077 | 0.297 | 0.295 |

Table 4 shows that the BiLSTM-LSTM algorithm and the BiLSTM-BiLSTM algorithm exhibit the highest increase ratios of 109.7% and 113.4%, respectively, compared to the BiLSTM algorithm across the five circular cross-section models. The LSTM-LSTM algorithm and LSTM-BiLSTM algorithm demonstrate effective increases of 27.5% and 25.4%, respectively. Within the lung cross-section model, the maximum increases for the BiLSTM-LSTM algorithm, BiLSTM-BiLSTM algorithm, LSTM-LSTM algorithm, and LSTM-BiLSTM algorithm are 11.7%, 7.7%, 14.6%, and 17.1%, respectively. Overall, the LSTM-LSTM algorithm and the LSTM-BiLSTM algorithm outperform the BiLSTM-LSTM algorithm and the BiLSTM-BiLSTM algorithm.

## 6 Conclusion

To enhance lung EIT image quality, a novel deep recurrent neural network leveraging LSTM and BiLSTM architectures was developed in this study. Four distinct algorithms were derived: the LSTM-LSTM algorithm, the LSTM-BiLSTM algorithm, the BiLSTM-LSTM algorithm, and the BiLSTM-BiLSTM algorithm. Conclusions were drawn from simulation results involving five sets of circular cross-section models and five sets of lung cross-section models.

In comparison to the LSTM algorithm, the LSTM-LSTM and LSTM-BILSTM algorithms exhibit maximum increases in correlation coefficients of 27.5% and 25.4%, respectively. Similarly, compared with the BiLSTM algorithm, the BILSTM-LSTM and BILSTM-BILSTM algorithms show increases of 11.7% and 13.4% in correlation coefficients. Overall, the quality of EIT images is observed.

In general, stacking neural networks effectively enhances the quality of the reconstructed images but also leads to an increase in the training time. Therefore, the stacked networks of neural networks should be done judiciously. In this research, the LSTM-LSTM algorithm and the LSTM-BiLSTM algorithm are generally superior to the BiLSTM-LSTM algorithm and the BiLSTM-BiLSTM algorithm in the performance of reconstructing images. Subsequent research will involve applying all algorithms to monitor lung function in practice. And the lung EIT images will be reconstructed in combination with CT images.

## Acknowledgements

This work was supported in part by the National Natural Science Foundation of China (52105564); The Key Science and Technology Plan Project of Jinhua City, China (2023-3-084); National Natural Science Foundation of China (52205075); Undergraduate Teaching Content and Curriculum System Reform Project, Shizhen College of Guizhou University of Traditional Chinese Medicine (SZXY2024005). (Corresponding author: Zhenzhong Song)